\documentclass[]{aastex631}
\shorttitle{}
\shortauthors{}
\graphicspath{{./}{figures/}}

\begin{document}

\title{A Potential New Mechanism for the Butterfly Diagram of the Solar Cycle:
Latitude-dependent Radial Flux Transport}

\correspondingauthor{Jie Jiang}
\email{jiejiang@buaa.edu.cn}

\author[0000-0002-2219-9829]{Zebin Zhang}
\affiliation{School of Space and Environment, Beihang University, Beijing, China}
\affiliation{Key Laboratory of Space Environment monitoring and Information Processing of MIIT, Beijing, China}

\author[0000-0001-5002-0577]{Jie Jiang}
\affiliation{School of Space and Environment, Beihang University, Beijing, China}
\affiliation{Key Laboratory of Space Environment monitoring and Information Processing of MIIT, Beijing, China}

\author{Haowei Zhang}
\affiliation{School of Space and Environment, Beihang University, Beijing, China}
\affiliation{Key Laboratory of Space Environment monitoring and Information Processing of MIIT, Beijing, China}



\begin{abstract}
The butterfly diagram of the solar cycle is the equatorward migration of the emergence latitudes of sunspots as the solar cycle evolves. Revealing the mechanism for the butterfly diagram is essential for understanding the solar and stellar dynamo. The equatorward meridional flow at the base of the convection zone (CZ) was believed to be responsible for the butterfly diagram. However, helioseismological studies indicate controversial forms of the flow, and even present poleward flow at the base of the CZ, which poses a big challenge to the widely accepted mechanism. This motivates us to propose a new mechanism in this study. Using a data-driven Babcock--Leighton--type dynamo model, we carry out numerical simulations to explore how the latitude-dependent radial flux transport affects the latitudinal migration of the toroidal field, under different meridional flow profiles. The results indicate that when the radial transport of the poloidal field at higher latitudes is sufficiently faster, the toroidal fields of a new cycle at higher latitudes are generated earlier than that at lower latitudes, and vice versa. Thus, the butterfly diagram is suggested to correspond to the time- and latitude-dependent regeneration of the toroidal field due to the latitude-dependent radial transport of the poloidal flux.
\end{abstract}



\section{Introduction}
The appearing latitudes of the sunspots move closer to the equator as the solar cycle evolves \citep{Maunder1904,Hathaway2015}, thus forming the butterfly-like pattern of the time--latitude diagram of sunspots. This butterfly diagram provides a strong constraint on the solar dynamo models \citep{Charbonneau2020}. Revealing its mechanism is at the center of understanding the solar magnetic cycle and even the stellar magnetism that maintains starspots.

The solar cycle is believed to be attributed to the large-scale dynamo process (the periodic conversion between the poloidal and toroidal magnetic fields mediated by the convective flow; \cite{Karak2014,Charbonneau2020}). The locations of the sunspots represent the distributions of the subsurface toroidal fields. The first attempt to explain the butterfly diagram in the framework of the dynamo theory is from \cite{Parker1955a}, who suggests that an equatorward dynamo wave of the toroidal field is responsible for the equatorward migration of the sunspots. Based on the Parker--Yoshimura rule \citep{Parker1955a,Yoshimura1975}, an equatorward propagation of the dynamo wave requires the rotation rate to increase with the depth for the positive $\alpha$-effect in the northern hemisphere, which is expected for the $\alpha$-effect in the bulk of the convection zone (CZ) due to cyclonic turbulence or the $\alpha$-effect near the surface due to the Babcock--Leighton (BL) mechanism. Meanwhile, helioseismology finds that the differential rotation rate in the overshoot layer, where the toroidal field is believed to be created and amplified \citep{Spiegel1980,Van1982,Choudhuri1990}, decreases with the depth at the latitudes of solar activity belts. Thus it brings into question the role of the dynamo wave played in the generation of the butterfly diagram.

The meridional flow provides another possibility responsible for the butterfly diagram. A poleward flow at the solar surface was found, implying an equatorward return flow within the Sun according to the conservation of mass. The single-cell meridional flow with an equatorward flow at the base of the CZ is widely assumed. Flux transport dynamo (FTD) models characterized by the key role played by such meridional flow \citep{Wang1991,Durney1995,Dikpati1999,Chatterjee2004} successfully reproduce not only the butterfly diagram, but also other basic features of the solar cycle. Most FTD models incorporate the BL source term, whose $\alpha$-effect is positive in the northern hemisphere based on observations. So the dynamo wave in such models is poleward at low latitudes. The return flow at the base of the CZ could overwhelm the poleward dynamo wave to generate the butterfly diagram \citep{Choudhuri1995}.

However, helioseismological results indicate that the meridional flow might not satisfy the requirement of the FTD models. \cite{Hathaway2012} find a shallow return flow 50 - 70 Mm beneath the surface. The helioseismic inversion from \cite{Zhao2013} also shows a shallow equatorward return flow located between 0.82 and 0.91 R$_\odot$. They further suggest a second meridional circulation cell below the shallower one. The second cell has a poleward flow at the base of the CZ. Results from \cite{Schad2013} indicate that the flow profile has a complex spatial structure consisting of multiple flow cells distributed in depth and latitude. Although recently \cite{Gizon2020} infer that the meridional flow is a deep single cell in each hemisphere, which is consistent with the prerequisite of FTD models, its final profile is still an open question. Magnetohydrodynamic simulations do not provide a conclusive result either. \cite{Hotta2022} show that the meridional flow has a double-cell profile. But most simulations \citep[e.g.,][]{Featherstone2015, Brun2017, Passos2017, Guerrero2019} show complex multicelled flow profiles in radial and latitudinal directions. And the profiles vary significantly from one model to another.


Abovementioned divergent profiles of meridional flow conflict with the requirement of FTD models. This situation poses a challenge to the understanding of the solar cycle in the framework of the FTD, especially to the reproducing of the butterfly diagram. \cite{Hazra2014} find that the butterfly diagram could be reproduced well in FTD models, providing the bottom flow at low latitudes is equatorward. Although this result relaxes the requirement on the specific profile of the meridional flow at the bulk of the CZ, the equatorward flow at its base is still a necessary condition, which cannot be guaranteed based on the current understanding of the flow. So new mechanisms for the butterfly diagram are needed. The latitudinal pumping \citep{Guerrero2008, Hazra2016} and the dynamo wave in the near-surface shear layer \citep{Pipin2011, Karak2016} were suggested as the candidates of the mechanism for the butterfly diagram.

In this paper, we will explore the effect of the latitude-dependent radial transport of the poloidal field on the latitudinal migration of the toroidal field at the base of the CZ, under different meridional flow profiles (deep single cell, shallow single cell, and double cell). We will demonstrate that the butterfly diagram could be attributed to the latitude-dependent radial transport of poloidal flux. 
The solar dynamo has obtained strong observational evidence to be of the BL type \citep{Espuig2010, Kitchatinov2011, Cameron2015, Jiao2021}, whose essence is that the poloidal field is regenerated by the evolution of the tilted sunspot groups on the solar surface. \cite{Jiang2013} construct an FTD model whose BL-type source term is based on the simulated solar surface large-scale field during cycles 15-21. Such kind of data-driven source term captures the essence of the BL mechanism and makes the surface evolution of poloidal flux as realistic as possible in the model. Meanwhile, the free parameters of the dynamo model are minimized. Such a model is helpful for us to concentrate on the transport process of the poloidal field and on the latitudinal distribution and time evolution of the toroidal field. Therefore, we will adopt the data-driven FTD model given by \cite{Jiang2013} to clarify the new mechanism for the butterfly diagram.

The paper is organized as follows. The data-driven BL-type dynamo model is described in Section 2. The simulations considering different meridional flows
are presented in Section 3.1. We demonstrate how the new mechanism operates in Section 3.2. We summarize our results in Section 4.

\section{MODEL} \label{sec:model}
The axisymmetric large-scale magnetic fields and flow
profiles are expressed in spherical coordinates as
\begin{equation}
  \textbf{B}(r,\theta,t)=B(r,\theta,t) \hat{\textbf{e}}_\phi+
  \nabla\times
  \left[A(r,\theta,t)\hat{\textbf{e}}_\phi\right],\label{eq:eq1}
\end{equation}
\begin{equation}
  \textbf{u}(r,\theta)=r \sin\theta \Omega(r,\theta) \hat
  {\textbf{e}}_\phi+\textbf{u}_{p}(r,\theta),\label{eq:eq2}
\end{equation}
where $B(r,\theta,t)\hat{\textbf{e}}_\phi$ and
$\nabla\times\left[A(r,\theta,t)\hat{\textbf{e}}_\phi\right]$
represent the toroidal and poloidal components of the magnetic
fields, respectively. The large-scale flow fields, i.e., the meridional flow and the angular velocity, are denoted by $\textbf{u}_{p}(r,\theta)$ and $\Omega(r,\theta)$, respectively.
In the kinematic framework, the BL-type dynamo equations
for the poloidal and toroidal magnetic fields
\citep[for reviews, see ][]{Charbonneau2020} are
expressed as
\begin{equation}
  \frac{\partial A}{\partial t}+\frac{1}{r\sin\theta}(\textbf{u}_{p}\cdot\nabla)(r\sin\theta A)
  =\eta\left(\nabla^{2}-\frac{1}{r^{2}\sin^{2}\theta}\right)A+S_{BL},\label{eq:eq3}
\end{equation}
\begin{equation}
  \frac{\partial B}{\partial t}+\frac{1}{r}\left[\frac{\partial(ru_{r}B)}
  {\partial r}+\frac{\partial(u_{\theta}B)}{\partial\theta}\right]=\eta\left(\nabla^{2}-\frac{1}
  {r^{2}\sin^{2}\theta}\right)B+r\sin\theta(\textbf{B}_{p}\cdot\nabla\Omega)+\frac{1}{r}\frac{d\eta}
  {dr}\frac{\partial(rB)}{\partial r},\label{eq:eq4}
\end{equation}
where $\eta$ represents the turbulent diffusivity, and $S_{\rm BL}$
is the BL-type source term for the poloidal field.

\begin{figure*}[t!]
  \gridline{\fig{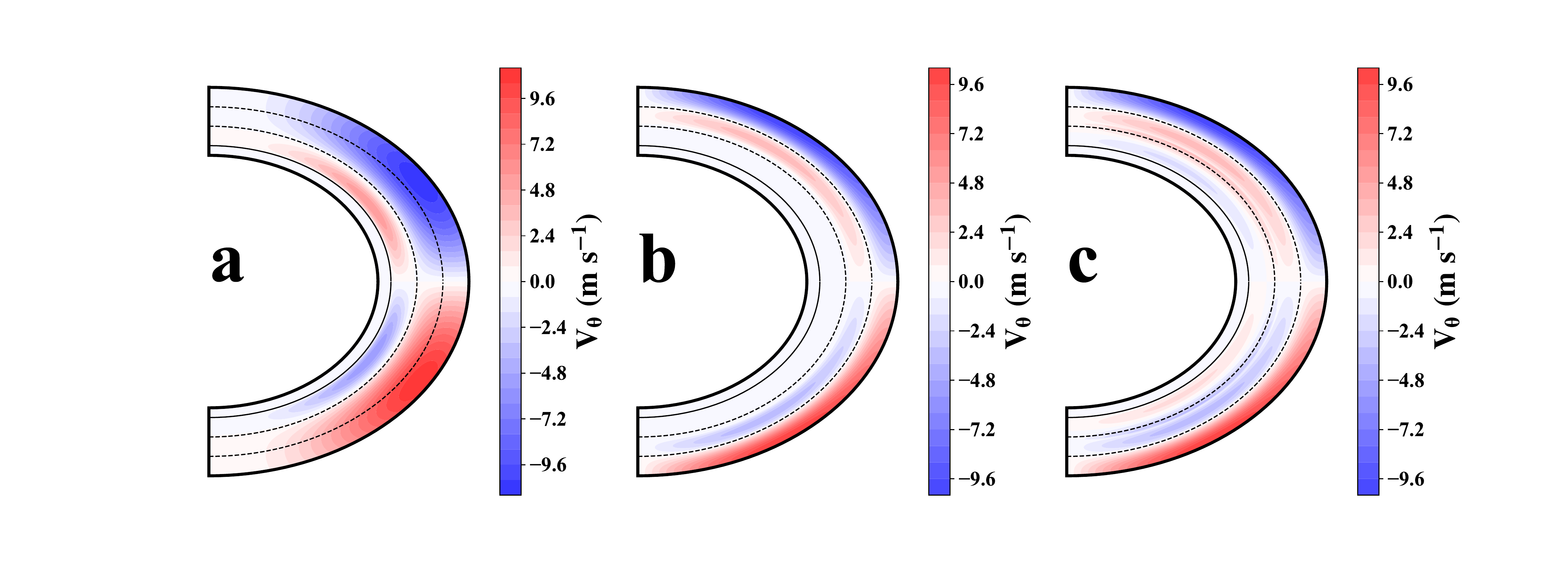}{1.0\textwidth}{}
  }
 \caption{Latitudinal velocities of the meridional flow used in the paper.
 (a) Deep single cell. (b) Shallow single cell.
 (c) Double cell.}
 \label{fig:Vq}
 \end{figure*}

\subsection{Data-driven source term}
The BL-type source term $S_{\rm BL}$ corresponds to the newly generated poloidal
field at the solar surface due to the decay
of tilted sunspot groups. The observed
magnetogram is a direct choice to construct this source
term. However, the available continuous magnetograms
just cover the recent four sunspot cycles.
Meanwhile, sunspot records have a much longer history.
Based on the historical sunspot records  \cite{Cameron2010} construct the BL-type
source term from 1913 to 1976 using surface flux
transport models. \cite{Jiang2013} apply this data-driven
source term into an FTD model. Here we follow \cite{Jiang2013}
to deal with the source term $S_{\rm BL}$.

\subsection{Meridional flow} \label{sec:flow}
In this study we consider three typical profiles adopted by the literature and will demonstrate that our results are not sensitive to the inner profiles of the flow. The three profiles are (i) a deep single-cell flow as same as the profile used by \cite{Dikpati2004} and \cite{Cameron2012}, i.e., Eqs. (11)-(16) of \cite{Cameron2012}; (ii) a shallow single-cell flow; and (iii) a double-cell flow as same as the profiles used by \cite{Hazra2014}. The latter two profiles correspond to Eqs. (10)-(12) of \cite{Hazra2014}. The only exception from them is that the amplitudes of the meridional flows used in this study are 11 m s$^{-1}$ at the surface.

Figure \ref{fig:Vq} shows the latitudinal velocities of the three profiles. For the deep single-cell flow the penetration depth is 0.7 $R_\odot$, and the return equatorward flow starts from 0.8 $R_\odot$. For the shallow single-cell flow, the penetration depth is 0.8 $R_\odot$, and the return equatorward flow starts from 0.9 $R_\odot$. For the double-cell flow, there are two cells stacked along the radial direction. The poleward flow locates between 0.7 and 0.78 $R_\odot$, and the equatorward flow locates between 0.8 and 0.9 $R_\odot$.

\subsection{Turbulent radial pumping} \label{sec:pumping}
Recently people are gradually aware of the important roles that radial pumping plays in the dynamo process \citep{Guerrero2008, Cameron2012, Kitchatinov2012, Karak2016, Kitchatinov2016, Karak2017, Zhang2022}. Magnetohydrodynamic simulations indicate
a latitude-dependent radial pumping, that is, the higher the
latitudes are, the stronger the pumping amplitude is \citep{Brandenburg1992, Ossendrijver2002,kapyla2006,Hotta2022}.
In this study, we adopt two kinds of radial pumping.

We adopt a latitude-independent pumping for the first case. That is
\begin{equation}
  \gamma(r) = \frac{\gamma_{cz}}{2}\left[1+erf\left(\frac{r-0.7R_\odot}{0.02R_\odot}\right)\right]
  +\frac{\gamma_s-\gamma_{cz}}{2}\left[1+erf\left(\frac{r-0.9R_\odot}{0.02R_\odot}\right)\right],\label{eq:eq7}
\end{equation}
where $\gamma_{cz}$ = -1.8 m s$^{-1}$ and $\gamma_{s}$ = -30.0 m s$^{-1}$, which are the pumping strength in the CZ and near surface, respectively. In the second case, we adopt a
latitude-dependent radial pumping by adding a
$|\cos\theta|$ term to Eq.(\ref{eq:eq7}). That is,
\begin{equation}
  \gamma_l(r,\theta) = \gamma(r)|\cos\theta|.\label{eq:eq6}
\end{equation}

The pumping effect could be viewed as a velocity field
adding to the meridional flow, so we replace
$\textbf{u}_p$ as $\textbf{u}_{p} + \gamma (r) \hat{e}_r$ or
$\textbf{u}_{p} + \gamma_l (r,\theta) \hat{e}_r$ in the dynamo Eqs.(\ref{eq:eq3}) and (\ref{eq:eq4}).

\subsection{Other ingredients}

Except the new profiles of the meridional flow and radial pumping listed in subsections \ref{sec:flow} and \ref{sec:pumping}, other ingredients of the model are kept the same as that in \cite{Jiang2013}. The ingredients include the profiles of differential rotation and diffusivity, boundary conditions, and initial conditions. The numerical simulation is computed using the code SURYA developed by A.R. Choudhuri and his colleagues \citep{Dikpati1994,Chatterjee2004}.

\begin{figure*}[t!]
	\gridline{\fig{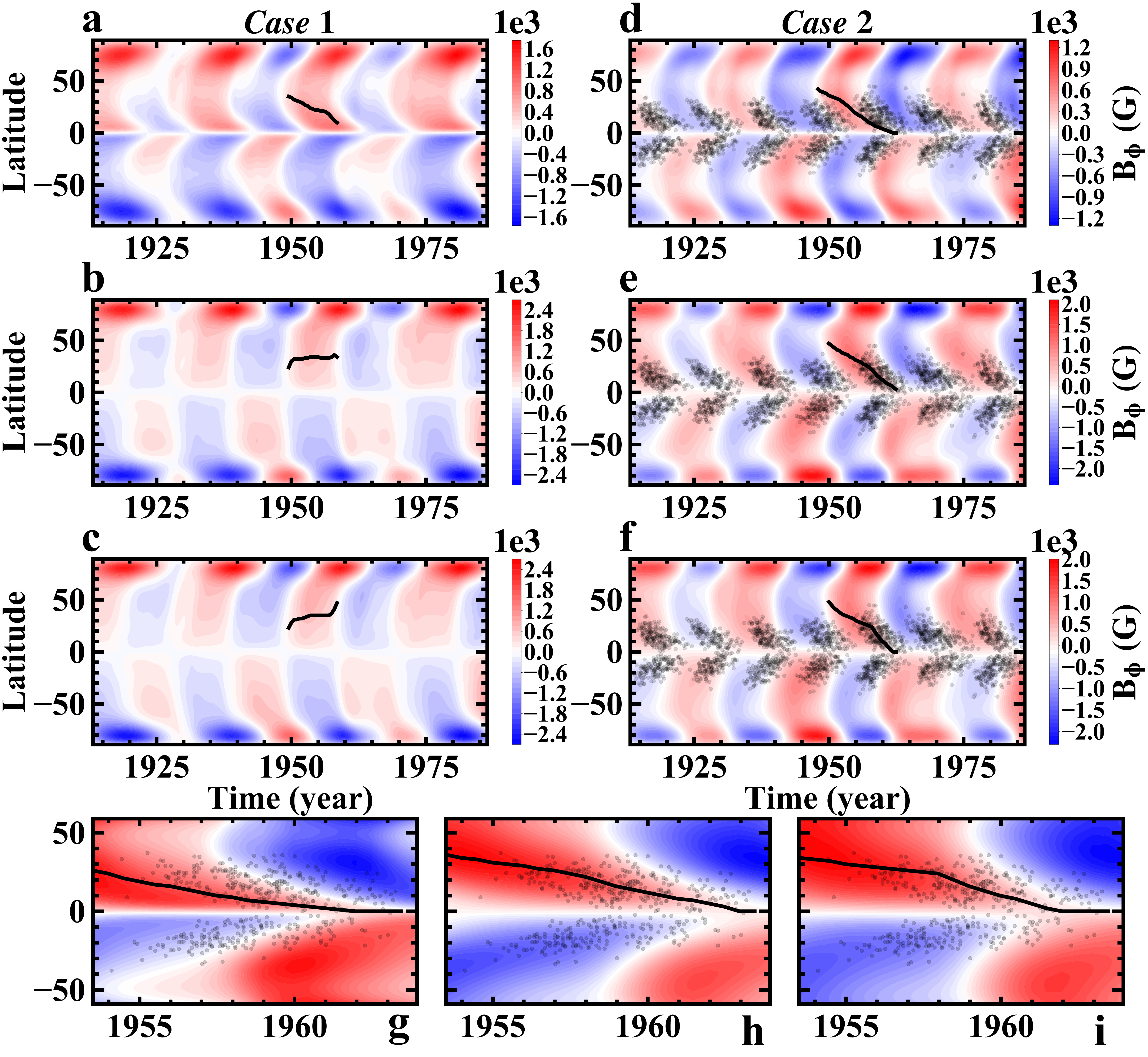}{0.9\textwidth}{}
	}
	\caption{Time--latitude diagrams of the toroidal field at the base of the CZ, $r$ = 0.715 $R_\odot$. Panels from top to bottom correspond to the three flow profiles shown in Figure \ref{fig:Vq}. Left panels correspond to Case 1 adopting the latitude-independent pumping. Right panels correspond to Case 2 adopting the latitude-dependent pumping. The butterfly wings during cycle 19 shown in panels (d) - (f) are enlarged, and the corresponding results are shown in panels (g) - (i). Black regions in panels (d) - (i) show the time--latitude evolution of observed sunspots. The black curve is the centroid of the toroidal flux between 0$^\circ$ and 55$^\circ$ latitudes, representing the migration of toroidal fields.}
	\label{fig:butbot}
\end{figure*}

\section{Results}
\label{sec:results}
\subsection{Effects of the latitude-dependent radial flux transport}

In this section, we will explore how the latitude-dependent radial transport of the poloidal flux affects the latitudinal migration of the toroidal fields at the base of the CZ, under different profiles of meridional flows. Figure \ref{fig:butbot} shows the time--latitude diagrams of the toroidal field at $r$ = 0.715 $R_\odot$. And panels from top to bottom correspond to the three flow profiles shown in Figure \ref{fig:Vq}. Figures \ref{fig:butbot}(a) - (c) correspond to Case 1 adopting the latitude-independent pumping, while Figures \ref{fig:butbot}(d) - (f) correspond to Case 2 adopting the latitude-dependent pumping. The periods of these solutions slightly change over cycles but are around 11 yr regardless of the profiles of meridional flow. This is because the cycle period is dominated by the inherent periodicity of the data-driven source term.

The simulation shown in Figure \ref{fig:butbot}(a) is similar to the reference case shown in Figure 4 of \cite{Jiang2013} since they have almost the same parameters. The basic features of the sunspot cycle are reproduced well. For instance, the toroidal fields at low latitudes propagate toward the equator and form the butterfly-like pattern, caused by the equatorward transport of toroidal flux under the effect of equatorward return meridional flow.

However, when the meridional flow is the shallow single or double cell, the solution does not show the butterfly-like pattern anymore. This is expected, as presented in previous studies \citep[e.g.,][]{Hazra2014}. For the solution shown in Figure \ref{fig:butbot}(b), there is no large-scale flow at the base of the CZ since the penetration depth of the meridional flow only extends to 0.8 $R_\odot$. At low latitudes, the poleward propagation of the toroidal field is caused by the dynamo wave. But in comparison with Figure 1(c) of \cite{Hazra2014}, the poleward propagational trend is weaker. This is because of the data-driven source term in our model. The observed surface poloidal field source includes some time difference at different latitudes, which delays the generation of the toroidal field at lower latitudes. This effect counteracts the effect of the dynamo wave, which causes the weaker trend. For the solution shown in Figure \ref{fig:butbot}(c), the meridional flow at the base of the CZ is poleward. This poleward flow works with the poleward dynamo wave. Hence a stronger poleward propagational tendency than that shown in Figure \ref{fig:butbot}(b) is obtained. The latitudinal migration of toroidal fields is illustrated by the black curves.

Then we replace the latitude-independent radial pumping with latitude-dependent radial pumping. The corresponding results are shown in Figures \ref{fig:butbot}(d) - (f), which are prominently different from Figures \ref{fig:butbot}(a) - (c). At latitudes lower than $\pm 50^\circ$, the toroidal fields all propagate toward the equator and form the solar-like butterfly diagrams. To explore if the butterfly diagrams shown in Figures \ref{fig:butbot}(d) - (f) match the real sunspot butterfly diagram, we superimpose the corresponding time evolution of sunspots’ latitudes onto the toroidal field. We zoom in on the butterfly wings shown in Figures \ref{fig:butbot}(d) - (f). The corresponding results are presented in Figures \ref{fig:butbot}(g) - (i). The toroidal field evolution (in red and blue) shown in Figures \ref{fig:butbot}(h) and (i) almost overlap with that of sunspots (black dots, based on USAF/NOAA data \footnote{http://solarcyclescience.com/activeregions.html}). And the propagation trend (represented by the black curve) is similar to the activity belt. Before the start of each sunspot cycle, there are toroidal fields at higher latitudes, e.g., around $\pm 50^\circ$. They could correspond to the ephemeral regions observed at the solar surface.

In Figure \ref{fig:butbot}(d) the maxima of the toroidal field and the maxima of the sunspot number show some time shift. See also the enlarged details in Figure \ref{fig:butbot} (g). This is because the two mechanisms, equatorward meridional flow and latitude-dependent radial flux transport, work together leading to the quick equatorward migration of the toroidal field. But it does not bring a significant decrease in the cycle period because the data-driven source term dominates the cycle period.

The solutions shown in Figure \ref{fig:butbot} all have a strong toroidal branch near the poles. It results from the strong radial shear in the tachocline at high latitudes, and is a common issue for FTD models that assume the toroidal fields are regenerated in the tachocline. People usually conjecture that this branch does not lead to sunspot emergence based on thin flux tube simulations \citep{Caligari1995, Fan2021}.


\begin{figure*}[t!]
	\gridline{\fig{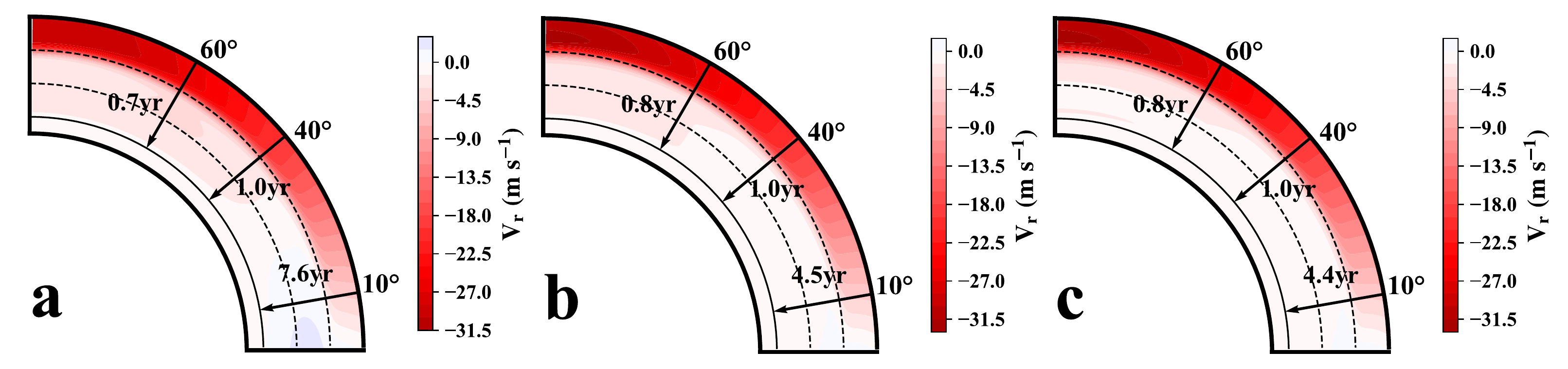}{0.8\textwidth}{}
	}
	\caption{Inward velocity distributions. Three panels correspond to that of Figure \ref{fig:Vq}
		with the addition of inward pumping of Eq.(\ref{eq:eq6}). The black arrows
		indicate the transport time of the poloidal fields from the surface to the base of the CZ, at latitudes 10$^\circ$, 40$^\circ$, and 60$^\circ$.}
	\label{fig:Velocity}
\end{figure*}

\subsection{Illustration of the potential new mechanism}

The simulations in the last subsection suggest that the latitude-dependent radial flux transport is a potential mechanism for the butterfly diagram. We use Figure \ref{fig:Velocity} to illustrate the velocity distributions of the inward transport of the poloidal field. The three panels correspond to the three profiles of the meridional flow given in Figure 1 with the addition of latitude-dependent pumping given by Eq.(\ref{eq:eq6}). We denote the time difference of the flux transport between $\pm 40^\circ$ and $\pm 10^\circ$ latitudes as $\Delta t$. Both the pumping and meridional flow contribute to $\Delta t$ for each panel. The different $\Delta t$ values among the three panels result from the different radial components of the meridional flow profiles. The deep single-cell flow has the largest $\Delta t$ because the flow at higher latitudes has a prominent quicker inward component than that at low latitudes as a whole. The other two flow profiles also have the high--low latitude difference although they are weaker. This indicates that our new mechanism for the butterfly diagram results from not only the latitude-dependent pumping but also the latitude-dependent radial component of meridional flow. The latitudinal migration pattern of the toroidal field depends on the competition between the meridional flow, the dynamo wave, and the latitudinal-dependent radial transport of the poloidal field. The time scale for the double-cell flow to transport the magnetic fields from $\pm 10^\circ$ to $\pm 40^\circ$ latitudes approximately is 20 yr. The time scale for the poleward dynamo wave is about 10 yr. Hence for different flow profiles, the required $\Delta t$ is different to overpower the poleward dynamo wave and poleward flow to generate the realistic butterfly-like pattern.

\begin{figure*}[t!]
	\gridline{\fig{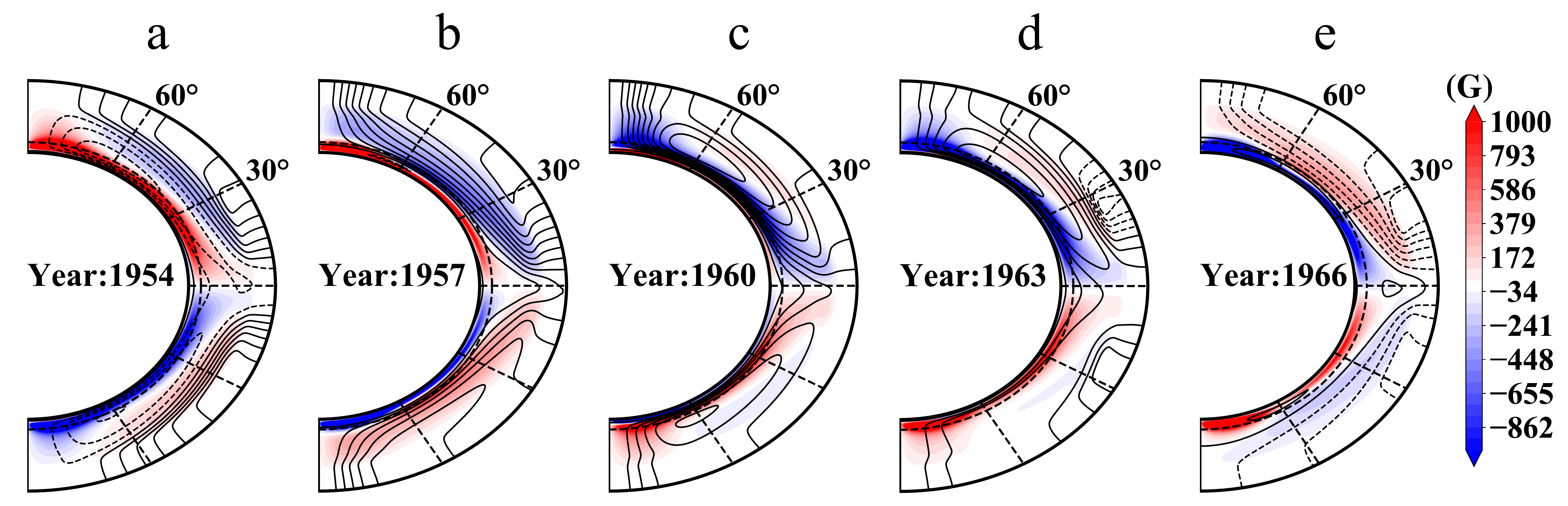}{1.0\textwidth}{}
	}
	\caption{Snapshots of the toroidal and poloidal fields for the solution shown
		in Figure \ref{fig:butbot}(e) (and also Figure \ref{fig:butbot}(h)).
		The red (blue) contours represent the positive (negative) toroidal field.
		Solid (dashed) lines represent clockwise (counterclockwise)
		poloidal field lines. The panels (a) - (e) correspond to the years of 1954, 1957, 1960, 1963, and 1966, respectively, covering the whole cycle 19.}
	\label{fig:ALL}
\end{figure*}

To further illustrate how the new mechanism for the butterfly diagram works, we present the snapshots of the toroidal and poloidal field evolution in Figure \ref{fig:ALL} corresponding to Figure \ref{fig:butbot}(h). Comparing with Figure 2 of \cite{Jiang2013} using the deep single-cell flow, we may see the notably different processes leading to the latitudinal migration of the toroidal field. In Figure 2 of \cite{Jiang2013}, the toroidal field of a new cycle first appears around $\pm 50^\circ$ latitudes due to the radial transport of the latitudinal component of the poloidal field $B_\theta$. Then the toroidal fields are equatorward transported under the effect of the equatorward return flow at the base of the CZ. In contrast, the transport of the toroidal field does not exist in Figure \ref{fig:ALL}. Toroidal fields at higher (lower) latitudes are generated earlier (later) than that at lower (higher) latitudes. Thus the butterfly pattern of the toroidal field evolution is generated. Figure \ref{fig:ALL}(a) clearly shows that around $\pm 60^\circ$ latitudes the toroidal field of the old cycle is first canceled, and the toroidal field of the new cycle is first generated due to the much quicker inward transport of the poloidal field. Figures \ref{fig:ALL}(b) - (e) show that the toroidal fields of the new cycle are gradually built up at lower and lower latitudes.

\section{Conclusion}
We have proposed a potential new mechanism for the butterfly diagram of the solar cycle. When latitude-dependent radial transport of poloidal flux exists, the toroidal field at higher latitudes could be regenerated earlier than that at lower latitudes. The time- and latitude-dependent regeneration of the toroidal field could cause the equatorward migration of the toroidal field. The new mechanism opens the possibility to generate the butterfly diagram under complex meridional flow patterns, even when the flow at the base of the CZ is poleward. The mechanism is distinct from the popular one, which results from the equatorward transport of the toroidal flux led by the meridional flow.

In the present study, both the meridional flow and radial pumping contribute to the latitude-dependent radial transport of the poloidal flux. The latitude-dependent radial pumping plays a major role. This does not mean that the latitude-dependent radial pumping is a necessity for the model. Other latitude-dependent radial transport mechanisms for the poloidal field could also help the model work. In addition, in the self-excited dynamo model operating in the bulk of CZ developed by \cite{Zhang2022}, they suggest another ingredient contributing to the time- and latitude-dependent regeneration of the toroidal field, the latitude-dependent latitudinal shear. It is the strongest at middle latitudes ($\sim55^\circ$), which makes the time for building up the toroidal field at middle latitudes shorter than that at lower latitudes. In the forthcoming work, we will systematically explore the factors contributing to the time- and latitude-dependent regeneration of the toroidal field for the butterfly diagram. The behavior in the self-excited dynamo models will also be investigated.

\begin{acknowledgments}
We sincerely thank the anonymous referee for the insightful comments and Robert Cameron for always valuable discussions. The research is supported by the National Natural Science Foundation of China (grant Nos. 12173005 and 11873023).
\end{acknowledgments}

\bibliography{sample631}{}

\begin{thebibliography}{}
\expandafter\ifx\csname natexlab\endcsname\relax\def\natexlab#1{#1}\fi
\providecommand{\url}[1]{\href{#1}{#1}}
\providecommand{\dodoi}[1]{doi:~\href{http://doi.org/#1}{\nolinkurl{#1}}}
\providecommand{\doeprint}[1]{\href{http://ascl.net/#1}{\nolinkurl{http://ascl.net/#1}}}
\providecommand{\doarXiv}[1]{\href{https://arxiv.org/abs/#1}{\nolinkurl{https://arxiv.org/abs/#1}}}

\bibitem[{{Brandenburg} {et~al.}(1992){Brandenburg}, {Moss}, \&
  {Tuominen}}]{Brandenburg1992}
{Brandenburg}, A., {Moss}, D., \& {Tuominen}, I. 1992, in Astronomical Society
  of the Pacific Conference Series, Vol.~27, The Solar Cycle, ed. K.~L.
  {Harvey}, 536

\bibitem[{{Brun} {et~al.}(2017){Brun}, {Strugarek}, {Varela}, {Matt},
  {Augustson}, {Emeriau}, {DoCao}, {Brown}, \& {Toomre}}]{Brun2017}
{Brun}, A.~S., {Strugarek}, A., {Varela}, J., {et~al.} 2017, \apj, 836, 192,
  \dodoi{10.3847/1538-4357/aa5c40}

\bibitem[{{Caligari} {et~al.}(1995){Caligari}, {Moreno-Insertis}, \&
  {Schussler}}]{Caligari1995}
{Caligari}, P., {Moreno-Insertis}, F., \& {Schussler}, M. 1995, \apj, 441, 886,
  \dodoi{10.1086/175410}

\bibitem[{{Cameron} \& {Sch{\"u}ssler}(2015)}]{Cameron2015}
{Cameron}, R., \& {Sch{\"u}ssler}, M. 2015, Science, 347, 1333,
  \dodoi{10.1126/science.1261470}

\bibitem[{{Cameron} {et~al.}(2010){Cameron}, {Jiang}, {Schmitt}, \&
  {Sch{\"u}ssler}}]{Cameron2010}
{Cameron}, R.~H., {Jiang}, J., {Schmitt}, D., \& {Sch{\"u}ssler}, M. 2010,
  \apj, 719, 264, \dodoi{10.1088/0004-637X/719/1/264}

\bibitem[{{Cameron} {et~al.}(2012){Cameron}, {Schmitt}, {Jiang}, \&
  {I{\c{s}}{\i}k}}]{Cameron2012}
{Cameron}, R.~H., {Schmitt}, D., {Jiang}, J., \& {I{\c{s}}{\i}k}, E. 2012,
  \aap, 542, A127, \dodoi{10.1051/0004-6361/201218906}

\bibitem[{{Charbonneau}(2020)}]{Charbonneau2020}
{Charbonneau}, P. 2020, Living Reviews in Solar Physics, 17, 4,
  \dodoi{10.1007/s41116-020-00025-6}

\bibitem[{{Chatterjee} {et~al.}(2004){Chatterjee}, {Nandy}, \&
  {Choudhuri}}]{Chatterjee2004}
{Chatterjee}, P., {Nandy}, D., \& {Choudhuri}, A.~R. 2004, \aap, 427, 1019,
  \dodoi{10.1051/0004-6361:20041199}

\bibitem[{Choudhuri(1990)}]{Choudhuri1990}
Choudhuri, A.~R. 1990, The Astrophysical Journal, 355, 733

\bibitem[{{Choudhuri} {et~al.}(1995){Choudhuri}, {Schussler}, \&
  {Dikpati}}]{Choudhuri1995}
{Choudhuri}, A.~R., {Schussler}, M., \& {Dikpati}, M. 1995, \aap, 303, L29

\bibitem[{{Dasi-Espuig} {et~al.}(2010){Dasi-Espuig}, {Solanki}, {Krivova},
  {Cameron}, \& {Pe{\~n}uela}}]{Espuig2010}
{Dasi-Espuig}, M., {Solanki}, S.~K., {Krivova}, N.~A., {Cameron}, R., \&
  {Pe{\~n}uela}, T. 2010, \aap, 518, A7, \dodoi{10.1051/0004-6361/201014301}

\bibitem[{{Dikpati} \& {Charbonneau}(1999)}]{Dikpati1999}
{Dikpati}, M., \& {Charbonneau}, P. 1999, \apj, 518, 508,
  \dodoi{10.1086/307269}

\bibitem[{{Dikpati} \& {Choudhuri}(1994)}]{Dikpati1994}
{Dikpati}, M., \& {Choudhuri}, A.~R. 1994, \aap, 291, 975

\bibitem[{{Dikpati} {et~al.}(2004){Dikpati}, {de Toma}, {Gilman}, {Arge}, \&
  {White}}]{Dikpati2004}
{Dikpati}, M., {de Toma}, G., {Gilman}, P.~A., {Arge}, C.~N., \& {White}, O.~R.
  2004, \apj, 601, 1136, \dodoi{10.1086/380508}

\bibitem[{{Durney}(1995)}]{Durney1995}
{Durney}, B.~R. 1995, \solphys, 160, 213, \dodoi{10.1007/BF00732805}

\bibitem[{{Fan}(2021)}]{Fan2021}
{Fan}, Y. 2021, Living Reviews in Solar Physics, 18, 5,
  \dodoi{10.1007/s41116-021-00031-2}

\bibitem[{{Featherstone} \& {Miesch}(2015)}]{Featherstone2015}
{Featherstone}, N.~A., \& {Miesch}, M.~S. 2015, \apj, 804, 67,
  \dodoi{10.1088/0004-637X/804/1/67}

\bibitem[{{Gizon} {et~al.}(2020){Gizon}, {Cameron}, {Pourabdian}, {Liang},
  {Fournier}, {Birch}, \& {Hanson}}]{Gizon2020}
{Gizon}, L., {Cameron}, R.~H., {Pourabdian}, M., {et~al.} 2020, Science, 368,
  1469

\bibitem[{{Guerrero} \& {de Gouveia Dal Pino}(2008)}]{Guerrero2008}
{Guerrero}, G., \& {de Gouveia Dal Pino}, E.~M. 2008, \aap, 485, 267,
  \dodoi{10.1051/0004-6361:200809351}

\bibitem[{{Guerrero} {et~al.}(2019){Guerrero}, {Zaire}, {Smolarkiewicz}, {de
  Gouveia Dal Pino}, {Kosovichev}, \& {Mansour}}]{Guerrero2019}
{Guerrero}, G., {Zaire}, B., {Smolarkiewicz}, P.~K., {et~al.} 2019, \apj, 880,
  6, \dodoi{10.3847/1538-4357/ab224a}

\bibitem[{{Hathaway}(2012)}]{Hathaway2012}
{Hathaway}, D.~H. 2012, \apj, 760, 84, \dodoi{10.1088/0004-637X/760/1/84}

\bibitem[{{Hathaway}(2015)}]{Hathaway2015}
---. 2015, Living Reviews in Solar Physics, 12, 4, \dodoi{10.1007/lrsp-2015-4}

\bibitem[{{Hazra} {et~al.}(2014){Hazra}, {Karak}, \& {Choudhuri}}]{Hazra2014}
{Hazra}, G., {Karak}, B.~B., \& {Choudhuri}, A.~R. 2014, \apj, 782, 93,
  \dodoi{10.1088/0004-637X/782/2/93}

\bibitem[{{Hazra} \& {Nandy}(2016)}]{Hazra2016}
{Hazra}, S., \& {Nandy}, D. 2016, \apj, 832, 9,
  \dodoi{10.3847/0004-637X/832/1/9}

\bibitem[{{Hotta} {et~al.}(2022){Hotta}, {Kusano}, \& {Shimada}}]{Hotta2022}
{Hotta}, H., {Kusano}, K., \& {Shimada}, R. 2022, \apj, 933, 199,
  \dodoi{10.3847/1538-4357/ac7395}

\bibitem[{{Jiang} {et~al.}(2013){Jiang}, {Cameron}, {Schmitt}, \&
  {I{\c{s}}{\i}k}}]{Jiang2013}
{Jiang}, J., {Cameron}, R.~H., {Schmitt}, D., \& {I{\c{s}}{\i}k}, E. 2013,
  \aap, 553, A128, \dodoi{10.1051/0004-6361/201321145}

\bibitem[{{Jiao} {et~al.}(2021){Jiao}, {Jiang}, \& {Wang}}]{Jiao2021}
{Jiao}, Q., {Jiang}, J., \& {Wang}, Z.-F. 2021, \aap, 653, A27,
  \dodoi{10.1051/0004-6361/202141215}

\bibitem[{{K{\"a}pyl{\"a}} {et~al.}(2006){K{\"a}pyl{\"a}}, {Korpi},
  {Ossendrijver}, \& {Stix}}]{kapyla2006}
{K{\"a}pyl{\"a}}, P.~J., {Korpi}, M.~J., {Ossendrijver}, M., \& {Stix}, M.
  2006, \aap, 455, 401, \dodoi{10.1051/0004-6361:20064972}

\bibitem[{{Karak} \& {Cameron}(2016)}]{Karak2016}
{Karak}, B.~B., \& {Cameron}, R. 2016, \apj, 832, 94,
  \dodoi{10.3847/0004-637X/832/1/94}

\bibitem[{{Karak} {et~al.}(2014){Karak}, {Jiang}, {Miesch}, {Charbonneau}, \&
  {Choudhuri}}]{Karak2014}
{Karak}, B.~B., {Jiang}, J., {Miesch}, M.~S., {Charbonneau}, P., \&
  {Choudhuri}, A.~R. 2014, \ssr, 186, 561, \dodoi{10.1007/s11214-014-0099-6}

\bibitem[{{Karak} \& {Miesch}(2017)}]{Karak2017}
{Karak}, B.~B., \& {Miesch}, M. 2017, \apj, 847, 69,
  \dodoi{10.3847/1538-4357/aa8636}

\bibitem[{{Kitchatinov} \& {Nepomnyashchikh}(2016)}]{Kitchatinov2016}
{Kitchatinov}, L.~L., \& {Nepomnyashchikh}, A.~A. 2016, Advances in Space
  Research, 58, 1554, \dodoi{10.1016/j.asr.2016.04.014}

\bibitem[{{Kitchatinov} \& {Olemskoy}(2011)}]{Kitchatinov2011}
{Kitchatinov}, L.~L., \& {Olemskoy}, S.~V. 2011, Astronomy Letters, 37, 656,
  \dodoi{10.1134/S0320010811080031}

\bibitem[{{Kitchatinov} \& {Olemskoy}(2012)}]{Kitchatinov2012}
---. 2012, \solphys, 276, 3, \dodoi{10.1007/s11207-011-9887-2}

\bibitem[{{Maunder}(1904)}]{Maunder1904}
{Maunder}, E.~W. 1904, \mnras, 64, 747, \dodoi{10.1093/mnras/64.8.747}

\bibitem[{{Ossendrijver} {et~al.}(2002){Ossendrijver}, {Stix}, {Brandenburg},
  \& {R{\"u}diger}}]{Ossendrijver2002}
{Ossendrijver}, M., {Stix}, M., {Brandenburg}, A., \& {R{\"u}diger}, G. 2002,
  \aap, 394, 735, \dodoi{10.1051/0004-6361:20021224}

\bibitem[{{Parker}(1955)}]{Parker1955a}
{Parker}, E.~N. 1955, \apj, 122, 293, \dodoi{10.1086/146087}

\bibitem[{{Passos} {et~al.}(2017){Passos}, {Miesch}, {Guerrero}, \&
  {Charbonneau}}]{Passos2017}
{Passos}, D., {Miesch}, M., {Guerrero}, G., \& {Charbonneau}, P. 2017, \aap,
  607, A120, \dodoi{10.1051/0004-6361/201730568}

\bibitem[{{Pipin} \& {Kosovichev}(2011)}]{Pipin2011}
{Pipin}, V.~V., \& {Kosovichev}, A.~G. 2011, \apjl, 727, L45,
  \dodoi{10.1088/2041-8205/727/2/L45}

\bibitem[{{Schad} {et~al.}(2013){Schad}, {Timmer}, \& {Roth}}]{Schad2013}
{Schad}, A., {Timmer}, J., \& {Roth}, M. 2013, \apjl, 778, L38,
  \dodoi{10.1088/2041-8205/778/2/L38}

\bibitem[{Spiegel \& Weiss(1980)}]{Spiegel1980}
Spiegel, E., \& Weiss, N. 1980, Nature, 287, 616

\bibitem[{Van~Ballegooijen(1982)}]{Van1982}
Van~Ballegooijen, A. 1982, Astronomy and Astrophysics, 113, 99

\bibitem[{{Wang} {et~al.}(1991){Wang}, {Sheeley}, \& {Nash}}]{Wang1991}
{Wang}, Y.~M., {Sheeley}, N.~R., J., \& {Nash}, A.~G. 1991, \apj, 383, 431,
  \dodoi{10.1086/170800}

\bibitem[{{Yoshimura}(1975)}]{Yoshimura1975}
{Yoshimura}, H. 1975, \apj, 201, 740, \dodoi{10.1086/153940}

\bibitem[{{Zhang} \& {Jiang}(2022)}]{Zhang2022}
{Zhang}, Z., \& {Jiang}, J. 2022, \apj, 930, 30,
  \dodoi{10.3847/1538-4357/ac6177}

\bibitem[{{Zhao} {et~al.}(2013){Zhao}, {Bogart}, {Kosovichev}, {Duvall}, \&
  {Hartlep}}]{Zhao2013}
{Zhao}, J., {Bogart}, R.~S., {Kosovichev}, A.~G., {Duvall}, T.~L., J., \&
  {Hartlep}, T. 2013, \apjl, 774, L29, \dodoi{10.1088/2041-8205/774/2/L29}

\end{thebibliography}
\bibliographystyle{aasjournal}



\end{document}